\input harvmac
\def \lc {light-cone\ }
\def \lr {\lref }
\def\bd {{{\del}_-}}
\def \pw {plane-wave\ }

\def \eq#1 {\eqno{(#1)}}

\def \a {\alpha}

\def \b {\beta}
\def \k {\kappa}

\def \om {\omega}

\def \p {\phi}
\def \ep {\epsilon}
\def \s {\sigma}

\def \r {\rho}

\def \G {\Gamma}
\def \l {\lambda}
\def \m {\mu}
\def \g {\gamma}
\def \n {\nu}

\def \e#1 {{\rm e}^{#1}}
\def \const {{\rm const }}

\def \ha { { 1\over 2 }}
\def \ov {\over}

\def \td {\tilde}

\def \m { \mu} \def \n { \nu} \def \k {\kappa} \def \l { \lambda}
\def \s {\sigma}

\def\mm { {\rm m}}

\def \rf {\refs}
\def\four{{1\over 4} }

\def \t {\theta} 
\def \eps {\epsilon}
\baselineskip8pt \Title{\vbox {\baselineskip 6pt{\hbox{Imperial/TP/01-02/029 }}{\hbox {
}}{\hbox {}}{\hbox{ }} {\hbox{ }}} } {\vbox{\centerline {
A class of exact pp-wave string models
 } \vskip4pt \centerline{    with interacting light-cone  gauge
actions
 } }} \vskip -20 true pt
\centerline { {Jorge G. Russo$^{a,b}$ 
\footnote {$^*$} {e-mail address: russo@df.uba.ar } 
and A.A. Tseytlin$^{c,d}$
\footnote{$^{\star}$}{\baselineskip8pt e-mail address:
tseytlin.1@osu.edu  } \footnote{$^{\dagger}$}{\baselineskip8pt
Also at Lebedev Physics Institute, Moscow.} }}
\medskip \smallskip
\centerline {\it {}$^a$ The Abdus Salam 
International Centre for Theoretical Physics, }
\smallskip
\centerline {\it I-34100 Trieste, Italy}
\medskip

\centerline {\it {}$^b$ Departamento de F\'\i sica, Universidad de
Buenos Aires, }
\smallskip
\centerline {\it Ciudad Universitaria and Conicet, Pab. I, 1428
Buenos Aires, Argentina}

\medskip

\centerline {\it {}$^c$ Theoretical Physics Group, Blackett
laboratory }
\smallskip
\centerline {\it Imperial College, London SW7 2BZ, U.K.}

\medskip

\centerline {\it {}$^d$ Smith Laboratory, Ohio State University }
\smallskip

\centerline {\it Columbus OH 43210-1106, USA }
\bigskip
\centerline {\bf Abstract}
\medskip
\baselineskip10pt \noindent

We find a general class of pp-wave string solutions
 with NS-NS $H_3$ or R-R $F_3$ field strengths, which are analogous to
solutions with  non-constant $F_5$ 
recently considered  by Maldacena and Maoz (hep-th/0207284).
We  show  that:  (i) all pp-wave solutions supported
by non-constant $H_3$ or $F_p$ fields are exact type II
superstring solutions to all orders in $\a'$;
(ii) the corresponding light-cone gauge Green-Schwarz
actions are non-linear in bosons  but 
always quadratic in fermions,  and describe UV finite 2-d
theories; (iii) the pp-wave backgrounds supported by non-constant
$F_3$ field do not have, in contrast to their  $F_5$-field counterparts,
``supernumerary'' supersymmetries and thus the associated
light-cone GS actions do not possess 2-d supersymmetry. 
We consider a specific example where the pp-wave $F_3$ background is
parametrized by an arbitrary holomorphic function of one complex bosonic coordinate. The corresponding GS action has the same bosonic part, similar Yukawa terms but twice as many interacting world-sheet fermions as  the
(2,2) supersymmetric model originating from the  analogous 
$F_5$ background. We also discuss the structure of massless
scalar vertex operators in the models related to $N=2$ super
sine-Gordon and $N=2$ super Liouville theories.

\medskip
\Date {August  2002}
\noblackbox
\baselineskip 16pt plus 2pt minus 2pt

\noblackbox \overfullrule=0pt

\lref\horts{G.~T.~Horowitz and A.~A.~Tseytlin, ``A New class of
exact solutions in string theory,'' Phys.\ Rev.\ D {\bf 51}, 2896
(1995) [hep-th/9409021].
}

\lref \rutone{ J.~G.~Russo and A.~A.~Tseytlin, ``Constant magnetic
field in closed string theory: an exactly solvable model,'' Nucl.\
Phys.\ B {\bf 448}, 293 (1995) [hep-th/9411099].
}

\lr\TT{ A.~A.~Tseytlin, ``Exact solutions of closed string
theory,'' Class.\ Quant.\ Grav.\ {\bf 12}, 2365 (1995)
[hep-th/9505052].
}

\lref\rrm {R.~R.~Metsaev,
``Type IIB Green-Schwarz superstring in plane wave Ramond-Ramond
background,''
Nucl.\ Phys.\ B {\bf 625}, 70 (2002)
[hep-th/0112044].
}

\lref \blau { M.~Blau, J.~Figueroa-O'Farrill, C.~Hull and
G.~Papadopoulos, ``A new maximally
 supersymmetric background ofIIB superstring theory,'' 
JHEP {\bf 0201}, 047 (2002) [hep-th/0110242]. 
 ``Penrose limits and maximal supersymmetry,
''Class.\ Quant.\ Grav.\ {\bf 19}, L87 (2002) 
[hep-th/0201081]. 
}

\lr \CG{C.~G.~Callan and Z.~Gan,
``Vertex Operators In Background Fields,''
Nucl.\ Phys.\ B {\bf 272}, 647 (1986).
}

\lr \mald {D.~Berenstein, J.~M.~Maldacena and H.~Nastase,
``Strings in flat space and pp waves from N = 4 super Yang Mills,''
JHEP {\bf 0204}, 013 (2002)
[hep-th/0202021].
}

\lr \mt { R.~R.~Metsaev and A.~A.~Tseytlin,
``Exactly solvable model of superstring in plane wave Ramond-Ramond
background,''
Phys.\ Rev.\ D {\bf 65}, 126004 (2002)
[hep-th/0202109].
}

\lr \all{ D.~Amati and C.~Klimcik, ``Strings In A Shock Wave
Background And Generation Of Curved Geometry From Flat Space
String Theory,'' Phys.\ Lett.\ B {\bf 210}, 92 (1988).
G.~T.~Horowitz and A.~R.~Steif, ``Space-Time Singularities In
String Theory,'' Phys.\ Rev.\ Lett.\ {\bf 64}, 260 (1990).
``Strings In Strong Gravitational Fields,'' Phys.\ Rev.\ D {\bf
42}, 1950 (1990).
G. Horowitz, in: {Proceedings of Strings '90}, College
Station, Texas, March 1990 (World Scientific,1991).
 }

\lr\metse{R.~R.~Metsaev and A.~A.~Tseytlin,
``Curvature cubed terms in  string theory effective actions,''
Phys.\ Lett.\ B {\bf 185}, 52 (1987).
}

\lr\mett{
R.~R.~Metsaev and A.~A.~Tseytlin,
``Order alpha-prime (two loop) equivalence of the string equations of
motion and the sigma model Weyl invariance conditions: dependence on the
dilaton and the antisymmetric tensor,''
Nucl.\ Phys.\ B {\bf 293}, 385 (1987).
}

\lref\HullVH{
C.~M.~Hull,
``Exact pp Wave Solutions Of 11-Dimensional Supergravity,''
Phys.\ Lett.\ B {\bf 139}, 39 (1984).
}

\lref\GuevenAD{
R.~Gueven,
``Plane Waves In Effective Field Theories Of Superstrings,''
Phys.\ Lett.\ B {\bf 191}, 275 (1987).
}

\lr\tpp{   A.~A.~Tseytlin,
``Finite sigma models and exact string solutions with Minkowski
signature metric,''
Phys.\ Rev.\ D {\bf 47}, 3421 (1993)
[hep-th/9211061].
``String vacuum backgrounds with covariantly constant null Killing
vector and 2-d quantum gravity,''
Nucl.\ Phys.\ B {\bf 390}, 153 (1993)
[hep-th/9209023].
``A Class of finite two-dimensional sigma models and string vacua,''
Phys.\ Lett.\ B {\bf 288}, 279 (1992)
[hep-th/9205058].
}

\lr\mam{J.~Maldacena and L.~Maoz,
``Strings on pp-waves and massive two dimensional field theories,''
hep-th/0207284.
}

\lr\RT{J.~G.~Russo and A.~A.~Tseytlin,
``On solvable models of type IIB superstring in 
NS-NS and R-R plane wave  backgrounds,''
JHEP {\bf 0204}, 021 (2002)
[hep-th/0202179].}

\lref\told{
A.~A.~Tseytlin,
``On singularities of spherically symmetric backgrounds in string
theory,''
Phys.\ Lett.\ B {\bf 363}, 223 (1995)
[hep-th/9509050].
}

\lref \kit{
C.~Klimcik and A.~A.~Tseytlin,
``Exact four-dimensional string solutions and Toda like sigma models
from 'null gauged' WZNW theories,''
Nucl.\ Phys.\ B {\bf 424}, 71 (1994)
[hep-th/9402120].
}
\lref\GW{
D.~J.~Gross and E.~Witten,
``Superstring Modifications Of Einstein's Equations,''
Nucl.\ Phys.\ B {\bf 277}, 1 (1986).
}

\lref\tsey{
A.~A.~Tseytlin,
``Ambiguity In The Effective Action In String Theories,''
Phys.\ Lett.\ B {\bf 176}, 92 (1986).
}
\lref \GSW{M.~B.~Green, J.~H.~Schwarz and E.~Witten,
``Superstring Theory. Vol. 1: Introduction,''
  Cambridge, UK: Univ. Pr. ( 1987) 469 P. 
} 

\lref\SSS{E.~Braaten, T.~L.~Curtright and C.~K.~Zachos,
``Torsion And Geometrostasis In Nonlinear Sigma Models,''
Nucl.\ Phys.\ B {\bf 260}, 630 (1985).
B.~E.~Fridling and A.~E.~van de Ven, 
``Renormalization Of Generalized Two-Dimensional Nonlinear
 Sigma Models,'' Nucl.\ Phys.\ B {\bf 268}, 719 (1986). 
C.~M.~Hull and P.~K.~Townsend,
``The Two Loop Beta Function For Sigma Models With Torsion,''
Phys.\ Lett.\ B {\bf 191}, 115 (1987).
R.~R.~Metsaev and A.~A.~Tseytlin,
``Two Loop Beta Function For The Generalized Bosonic Sigma Model,''
Phys.\ Lett.\ B {\bf 191}, 354 (1987).
D.~Zanon,
``Two Loop Beta Functions And Low-Energy String Effective Action For The
Two-Dimensional Bosonic Nonlinear Sigma Model With A Wess-Zumino-Witten
Term,''
Phys.\ Lett.\ B {\bf 191}, 363 (1987).
  D.~R.~Jones,
``Two Loop Renormalization Of D = 2 Sigma Models With Torsion,''
Phys.\ Lett.\ B {\bf 192}, 391 (1987).
S.~V.~Ketov, A.~A.~Deriglazov and Y.~S.~Prager,
``Three Loop Beta Function For The Two-Dimensional Nonlinear Sigma Model
With A Wess-Zumino-Witten Term,''
Nucl.\ Phys.\ B {\bf 332}, 447 (1990).
``Four Loop Divergences Of The Two-Dimensional (1,1) Supersymmetric
Nonlinear Sigma Model With A Wess-Zumino-Witten Term,''
Nucl.\ Phys.\ B {\bf 359}, 498 (1991).
}
\lr \muk {  S.~Mukhi,
``Finiteness Of Nonlinear Sigma Models With Parallelizing Torsion,''
Phys.\ Lett.\ B {\bf 162}, 345 (1985).
}

\lr\kuta{
D.~Kutasov and N.~Seiberg,
``Noncritical Superstrings,''
Phys.\ Lett.\ B {\bf 251}, 67 (1990).
}

\lr\horkap{K.~Hori and A.~Kapustin,
``Duality of the fermionic 2d black hole and N = 2 Liouville theory as
mirror symmetry,''
JHEP {\bf 0108}, 045 (2001)
[hep-th/0104202].
}

\lr\toi{A.~A.~Tseytlin, 
``On field redefinitions and exact solutions in string theory,'' 
Phys.\ Lett.\ B {\bf 317}, 559 (1993) [hep-th/9308042].
 } 
\lref\howe{
M.~T.~Grisaru, P.~S.~Howe, L.~Mezincescu, B.~Nilsson and P.~K.~Townsend,
``N=2 Superstrings In A Supergravity Background,''
Phys.\ Lett.\ B {\bf 162}, 116 (1985).
}

\lref\mtt{
R.~R.~Metsaev and A.~A.~Tseytlin,
``Superstring action in $AdS_5 \times S^5$: kappa-symmetry light cone gauge,''
Phys.\ Rev.\ D {\bf 63}, 046002 (2001)
[hep-th/0007036].
R.~R.~Metsaev, C.~B.~Thorn and A.~A.~Tseytlin,
``Light-cone superstring in AdS space-time,''
Nucl.\ Phys.\ B {\bf 596}, 151 (2001)
[hep-th/0009171].
}

\lr \john{J.~H.~Schwarz,
``Covariant Field Equations Of Chiral N=2 D = 10 Supergravity,''
Nucl.\ Phys.\ B {\bf 226}, 269 (1983).
}

\lr\cvet{M.~Cvetic, H.~Lu and C.~N.~Pope, 
``M-theory pp-waves, Penrose limits and supernumerary supersymmetries,'' 
hep-th/0203229.
``Penrose limits, pp-waves and deformed M2-branes,''
hep-th/0203082.
}

\lr\mab{N. Berkovits and J. Maldacena,
``N=2 Superconformal 
Description of Superstring in Ramond-Ramond Plane Wave Backgrounds",
[hep-th/0208092].
}

\lr\hsu{S.~J.~Hyun and H.~J.~Shin, 
``N=(4,4) Type IIA String Theory on PP-Wave Background,'' 
hep-th/0208074. 
 }

\lref \hhh{ V.~Sahakian, ``Strings in Ramond-Ramond backgrounds,'' hep-th/0112063. 
}

\lref \nw { C.~R.~Nappi and E.~Witten, ``A WZW model based on a
nonsemisimple group,'' Phys.\ Rev.\ Lett.\ {\bf 71}, 3751 (1993)
[hep-th/9310112].
}

\lref\cvst{
M.~Cvetic, H.~Lu, C.~N.~Pope and K.~S.~Stelle,
``T-duality in the Green-Schwarz formalism, and the massless/massive IIA  duality
map,''
Nucl.\ Phys.\ B {\bf 573}, 149 (2000)
[hep-th/9907202].
}

\lr \MMM{R.~R.~Metsaev and A.~A.~Tseytlin,
``Type IIB superstring action in AdS(5) x S(5) background,''
Nucl.\ Phys.\ B {\bf 533}, 109 (1998)
[hep-th/9805028].
}

\lr \koba{
K.~I.~Kobayashi and T.~Uematsu,
``N=2 supersymmetric Sine-Gordon theory and conservation laws,''
Phys.\ Lett.\ B {\bf 264}, 107 (1991).
}

\lr\iva{
E.~A.~Ivanov and S.~O.~Krivonos,
``U(1) Supersymmetric Extension Of The Liouville Equation,''
Lett.\ Math.\ Phys.\  {\bf 7}, 523 (1983)
[Erratum-ibid.\  {\bf 8}, 345 (1984)].
}

\newsec{Introduction}
Finding  new   string models 
with Minkowski signature  which are exact in $\a'$ and 
whose spectrum  can be explicitly determined 
is of great interest from the point of view 
of  better understanding of string theory 
in curved (cosmological, black-hole, etc.) backgrounds. 

One class  of such  models has  metric admitting a  covariantly-constant
null Killing vector.
Many of such  pp-wave  backgrounds 
with metric and other fields  having ``null'' structure 
\rf{\all,\horts} 
are  simple examples of $\a'$-exact solutions of string theory (see
\TT\ for a review). 

Some of these backgrounds 
give  exactly solvable (in terms of free oscillators
in \lc gauge)  non-compact curved-space superstring
models, for which  one can find the string spectrum and
compute some simplest ``observables''
(partition function, some correlation functions, etc.) 
 in much the same way as in
flat space \rutone. 
As was recently realized
 \refs{\rrm,\mald,\mt,\RT}, this
solvability property applies to string models corresponding not
only to the NS-NS but also to certain R-R \rf{\blau} \pw
backgrounds.

It is then  natural to look  for  more complicated cases 
where  the \lc action  is no longer  quadratic  but  
may  be  integrable, so that the corresponding  string spectrum 
may still  be possible  to 
determine. 
An interesting  example  of such models 
(representing  a pp-wave metric supported by 
a particular non-constant  R-R  5-form background) 
 was recently proposed  in  \mam.

In this paper
 we first  study    a general class of NS-NS 
models based on pp-wave metric and ``null'' 3-form  $H_3$ 
background depending on arbitrary  harmonic functions $b_m(x)$
($\del^2 b_m=0, \ m=1,2,...$) 
of the transverse  coordinates $x_i$.
The special case  when $b_m$ are linear in $x_i$ 
corresponds to homogeneous \pw backgrounds with constant  $H_3$ 
field  (these are, in fact, WZW models \nw). 
Another  special case which we consider
is 
 where  $b_m$ are chosen to be holomorphic functions 
of complex combinations ($z_1= x_1 + i x_2,...$)  of coordinates $x_i$.
The    R-R  counterpart of this background 
 -- with $H_3$ replaced by the R-R
3-form 
$F_3$ --  is the direct analog  of the $F_5$-solution   of  \mam.

We shall demonstrate   that all such  backgrounds 
 (both the NS-NS and the R-R ones, including the  one  of \mam)  are
exact superstring solutions to all orders in $\a'$: 
 all corrections to the leading-order field 
equations should vanish in a  natural scheme. 
A direct world-sheet  proof of  exactness of the non-constant 
$F_5$-background of \mam\ (and certain other 
supersymmetric pp-wave backgrounds) 
was recently given in \mab.
The space-time argument presented here is universal: it
applies to general pp-wave string models with a non-constant $F_p$
field.

The
   NS-NS models   describe  a new class
 of exact string solutions, 
different from the 
chiral null models  \horts\
   which are  also exact in $\a'$ but in which   
 the form of the $B_2$
background is 
   correlated with  the off-diagonal terms in the metric.

{}For the pp-wave  background with  a R-R $F_3$ field,  
the holomorphic 
functions  parametrizing the 
background 
can be chosen so that the bosonic part of the
light-cone gauge  GS Lagrangian  is that of an integrable 
(e.g., super sine-Gordon or super Liouville) model. 
The same pp-wave metric can be supported  by different 
R-R field strengths, leading to solutions 
with different amounts (and types of)  of supersymmetry.  
In contrast to the $F_5$ models  of \mam, 
here the fermionic GS extension  of the same 
bosonic  \lc  action, while still representing   a UV finite  2-d 
theory, is not (2,2) supersymmetric: in the $F_3$ case 
there are twice  as many  fermions and  
the  coefficient  
of the Yukawa interaction  term  is smaller 
 by  factor of 
$\sqrt {2}$.  The corresponding string background preserves
8 space-time 
supersymmetries, whose  role in the light-cone gauge is to 
imply the existence of the same  number of massless 
fermions  which are decoupled from the rest of the fields.

There are several  reasons which make these non-supersymmetric 1+1 dimensional
models  worth of  further  study.  In particular,
like their (2,2) supersymmetric counterparts, 
they may turn out to be integrable.
This property would allow one
to determine the corresponding string spectra.
Another  interesting  question is 
whether the   NS-NS models 
describing similar inhomogeneos pp-wave backgrounds,  which have an advantage 
of their  covariant-gauge action  (having as usual (1,1) world-sheet 
supersymmetry)
being explicitly known, 
may also be integrable. 
 
\bigskip

 Before proceeding to the main  topic of this paper, let us 
recall some  other previously known 
 embeddings of  interacting non-conformal 
 2-d theories  (with  $d$-dimensional  Euclidean-signature 
target space)  into  $\a'$-exact conformal sigma models  (with 
$2+d$ dimensional  Minkowski-signature  target space).

Given a generic non-conformal 
  sigma model with  a  curved Euclidean $d$-dimensional space  
 $
 ds^2_d = g_{ij}(x)  dx^i dx^j 
 $
as a target space and 
 with the  RG  beta-function 
   $\beta_{ij} (g)  = R_{ij} + ...$, 
 one can construct \tpp\  a  special  Weyl-invariant sigma-model 
 (i.e. a string solution) 
  with the following 
 $2+d$ dimensional Minkowski-signature   metric  and the
dilaton 
 \eqn\asd{
 ds^2_{2+d}  = dudv + g_{ij}(x,u)  dx^i dx^j  \ , \ \ \  \ \ \ \ \ 
 \phi(x,u,v) =  v + \td \phi(u,x)   \ . } 
Here  $g_{ij}(x,u)$   is subject to  the  
 1-st order differential equation
\eqn\sxa{
{\del\over \del u}  g_{ij}(x,u) =   \beta_{ij} ( g (x,u))\ , }
 which is nothing but the RG equation in 
 the ``transverse''
$d$-dimensional theory, with $u$ interpreted as a logarithm of the 
2-d UV scale. 
A  class  of  exact string 
 solutions  \tpp\  is found, in particular, 
 in the case when the
transverse 
model is a  (2,2) supersymmetric  Einstein-Kahler  sigma model
(for which the $\beta_{ij}$-function 
  has only the one-loop term). 
 The simplest $2+d=4$
dimensional 
example 
is provided by the $O(3)$ sigma model with $g_{ij}(x)$ being the
$S^2$-metric. Here 
    $ g_{ij}(x,u) = u g_{ij}$  and  thus 
\eqn\ses{
 ds^2_{2+ d}  =  
dudv \  + \   u\  (d\theta^2 + \sin^2 \theta\ d\varphi^2)  \ ,\ \ \ \ \
\ \ \ \ 
\phi(x,u,v) =  v + \four  \ln\ u   \ . } 
The key feature of these solutions  is that the 
dilaton is non-trivial, and contains, 
in particular,  the term linear in $v$.
This means that in the 
 \lc gauge description, the  non-conformal 
transverse  theory  should 
   be supplemented  by  the   definition of 
the stress tensor following from   the  original 
``$2+d$-dimensional'' covariant action. 

 In another example we would like to mention  one 
 promotes  an interacting   Toda-type
2-d QFT  to a  Minkowski-signature string solution in $2+d$ dimensions
by using  the following  pp-wave background with linear dilaton \kit\
($\r_i$=const,\ $i=1,...,d$)
\eqn\drem{
ds^2_{2+ d}=dudv + K(x) du^2 + dx_i dx_i    \ ,\ \ \ \ \ \ \ \ \ 
\phi= \r_i x_i  \ .  
}
The string equations are then  satisfied to all orders provided
\eqn\solw{
\del_i \del_i  K - 2 \r_i  \del_i K =0  \ ,  \ \ \ \ \ \ \ \ \
\r_i\r_i=  
{\textstyle  {1 \ov 4 \a'} } (8 - d) \ .  } 
The solution for $K$ can be chosen as a sum of exponents:
$ K = \sum_n c_n e^{ \a_{ni} x_i}$,  where for each $n$  we should have 
$\ \a_{ni} \a_{ni} = 2 \r_i \a_{ni}$. 
A particular case  is provided by   the Toda model potential,
i.e. we  end up with   the Toda model  as the  \lc gauge
theory.\foot{Explicitly,
the Toda model corresponds to
   the case  when  $\ \a_i$
are simple roots of the Lie   algebra of a maximally non-compact
real Lie group $G$ of rank $N=d-2 $
 and $\r_i= \ha
\sum_n  \a_{ni}$ is
 half  of the sum of all positive roots. }
The  sigma model corresponding  to  \drem\  is T-dual (in $y=u-v$
coordinate)  to a 
sigma model associated with a
particular $G/H$  ``null-gauged''   WZW model \kit.\foot{There 
  $H$ is  a
nilpotent subgroup of $G$
generated by $N-1$ simple roots (this condition on $H$
 is needed to get  a model with only one time-like  direction).
  The flat transverse coordinates $x^i$ correspond to   the Cartan
subalgebra
generators.}
\bigskip

The rest of the   paper  is organized as follows.

In section 2 we present a class of pp-wave solutions with 
non-constant $H_3$ form
and constant dilaton. We argue that all similar pp-wave 
 backgrounds supported
by  NS-NS or R-R fields (including the ones of \mam) 
 should be   exact solutions of superstring theory.
We also determine conditions for residual supersymmetry of the 
$H_3$-background. 
In section 3 we discuss a subclass of 
NS-NS solutions which are parametrized
by holomorphic
functions, determining the fractions of  supersymmetry  they 
preserve.  
In section 4 we write down 
the (1,1) supersymmetric RNS 
sigma model Lagrangians for these    backgrounds.

In section 5.1  we  find   the \lc gauge GS actions
associated with   pp-wave backgrounds 
with  non-constant   $F_p$ fields.
We first show that the corresponding  interacting 2-d  actions  are 
always  quadratic in GS fermions and thus are easy 
to write down. We  then  prove (in section 5.2) 
that these \lc actions 
define   UV finite 
 2-d  theories, in agreement with the general 
$\a'$-exactness  argument of section 2.2.

In section 5.3 we  consider explicitly the  \lc GS  action for 
the pp-wave   R-R  $F_3$ background  which is ``S-dual'' to the NS-NS  background of section 2.1.
This background
does not admit ``supernumerary''  Killing spinors
(i.e. the ones which are not annihilated by the \lc gauge condition \cvet), 
and thus 
the associated \lc GS action  does not have ``accidental''
linearly realized  2-d supersymmetry. 
This is in contrast to the case of  $F_5$-background 
of \mam\ where a similar  model 
(with the same bosonic part)  is (2,2) supersymmetric. 

Finally, in section 6, we consider  two examples 
of  backgrounds parametrized by a holomorphic function,  which are 
related to integrable models. We present  a discussion of  the structure 
of the associated  massless vertex operators,
which  applies  also to the model of \mam .

\newsec{Supersymmetric pp-wave model with  non-constant 
 NS-NS    3-form background}

\subsec{\bf The form of the string solution}

Let us consider the following ansatz for the metric and NS-NS 2-form
potential 
in 10-dimensional superstring theory 
\eqn\gme{
ds^2=dudv + K(x)du^2+dx_i^2+dy_m^2\ , \ \ \ \ \ 
\ \ i=1,...,d\ , \ 
\ \  m=d+1,..,8\ , }
\eqn\uga{
B_2= b_m (x)\ du\wedge dy_m
\ , \  \ \ \ \ \  H_3 = \del_i  b_m(x)\ dx_i\wedge  du\wedge dy_m\ . 
}
We split the 8 transverse  coordinates into  two groups -- $x_i$ and $y_m$,
with the functions $K$ and $b_m$  depending only on $x_i$. 
We set the dilaton to be constant, but there is a straightforward 
generalization to the case  when $\p= \r_i x_i + \td \phi(u)$
(which makes possible  to  include
the background  \drem,\solw\ as a special case).
Another obvious generalization is to  allow  $K$ and $b_m$ 
to depend on $u$. 

In view of the structure of the background
(in particular, the fact that 
 the non-zero part of curvature of the metric \gme\ is $R_{uiuj} 
= - \ha \del_i \del_j K$)
the non-trivial components  of the
 leading-order string field  equations  are 
 $D_i H_{i ku}=0$ and 
$R_{uu} - {1 \ov 4} H_{ukl} H_u^{\ kl} =0$. They imply 
(we always assume summation over $i$ and $m$)\foot{In the case of 
a non-trivial dilaton $\p= \r_i x_i$
 these equations become (cf. \solw):
$(\del_i - 2 \r_i) \del_i b_m =0,  \ \   
(\del_i - 2 \r_i) \del_i  K + \del_i b_m  \del_i b_m   =0.$
}
\eqn\asw{
\del_i \del_i b_m =0\ , \ \ \ \ \ \  \ \   
\del_i \del_i  K + \del_i b_m  \del_i b_m   =0\ .
}
Thus 
  $b_m$ can be  any set of  harmonic functions of $x_i$, while 
the general solution for $K$  can be written as 
\eqn\gensl{
K=-\ha b_m b_m  + K_0\ ,\ \ \ \ \  \ \  \del_i\del_i K_0=0\ . 
}
There are several special cases.
For  $b_m =0$  we recover the 
standard pp-wave solution with $K=K_0(x)$ being a harmonic function.
For  linear $b_m$, i.e. constant $H_3$,  and $K_0=0$ we get 
\eqn\lii{
b_m= f_{mi} x_i \ ,\ \ \ \ \ 
H_3= f_{mi}  dx_i\wedge  du\wedge dy_m \  ,  \ \ \ \
 K= -\ha w_{ij} x_i x_j \ , \ \ \ 
  w_{ij} \equiv   f_{mi} f_{mj} \ , }
where $w_{ij}$  (which is  the mass matrix  for $x_i$ in  the \lc  gauge
action)
is  non-negative.
 The corresponding models can be interpreted 
as WZW  theories for non-semisimple groups \nw\  (see also
 \rf{\rutone,\RT}).\foot{In 
this case the generalized curvature $\hat R$  discussed in the next 
subsection  vanishes, as  it should be for a 
parallelizable space.}

Another regular solution  of the equation for 
$b_m$  which does not require an  introduction of singularities or 
sources  is  given by a  quadratic ``traceless''  form:\foot{In 
 what follows we  set the solution $K_0$   of the
homogeneous equation in \gensl\ to zero.}
\eqn\iqual{
b_m = c_{mij} x_i x_j 
\ ,\  \ \ \ \   c_{m ii}=0\  ,  \  \ \ \ \  
K = - \ha d_{ijkl} x_i x_j x_k x_l\ , \ \ \ \ 
 d_{ijkl} \equiv  c_{mij} c_{mkl} \ .  }
This gives a quartic non-negative potential in the light-cone
gauge.\foot{
The  potential will  always  have  flat directions.
{}For example, one can choose  $b_m = c_m (x^2_1- x^2_2)$,  so that
$K= - \ha  c^2 (x^2_1- x^2_2)^2$.} 
An example of a singular solution (which needs to be supported by a
delta-function source) 
is  
$b_m = {Q_m\ov x^{d-2}}$   and   $K=-{  Q_m^2\ov 2x^{2d-4}}$.
\foot{Note that such $K$  decays  twice as fast  as the
harmonic function $K=K_0$ of the standard pp-wave 
solution with $H_3=0$.} 

The  Laplace equation for $b_m$ in \asw\  can be solved  also 
by choosing $b_m$ to be, e.g.,  real  part 
of holomorphic functions of complex combinations of coordinates 
$z_1=x_1 + i x_2, $ etc. 
The corresponding string models  parametrized by holomorphic functions 
  will be discussed below in section 3. 
Their  R-R counterparts (with $H_3$ replaced by  ``S-dual'' $F_3$
background)
are direct analogs of the  pp-wave solution \mam\  supported by 
a non-constant 5-form
background.
Such  solutions  will, in general,  be singular  at
some points,
i.e. they  will  require  extra
assumptions 
about sources or string-theory resolution  of the singularities
(cf. \told). 

\medskip

Finally, let us mention   that lifts of  the above solutions to 11
dimensions
belong to a  class of $D=11$ pp-wave backgrounds 
 first  considered  in \HullVH\ (a  general discussion of pp-waves in
$D=10$ supergravity 
appeared also  in \GuevenAD). 
There  the 4-form background was  chosen as $F_4 = f_3\wedge   du $
with 3-form $f_3$ being  any closed and co-closed form 
 and  $K$-function in the metric  satisfying   $\del^2 K = |f_3|^2$. 
The present  NS-NS solutions correspond to a specific choice of 
 $f_3= \del_i b_m(x) \ dx_i \wedge dy_m \wedge  dx_{11}$.
  The focus  of the present paper is not 
on  studying   general  types 
of pp-wave solutions of supergravity as such, but on identifying 
backgrounds 
which are exact in $\a'$, and  which  lead to new exactly solvable 
superstring models.

\bigskip

\subsec{\bf Exactness in $\a'$ }
We shall now  show that the above background represents  an exact
string solution, i.e. there exists a scheme in which  it is 
not modified to all orders in $\a'$. 
We shall first  give a general argument  based  on the structure of 
low-energy effective action  in closed  superstring theory. 
This argument will apply not only to the NS-NS background \gme,\uga\ but
also 
to any similar R-R  background  with the  pp-wave 
metric \gme\  supported  by a p-form  field strength  which has 
a ``null''
form 
\eqn\spe{
F_p = f_{k_1  ... k_{p-1}}(x)\  dx^{k_1} \wedge ... \wedge dx^{k_{p-1}}
\wedge  du \ .}
In particular, this will also  demonstrate  the 
exactness of the pp-wave
background 
considered in \mam. 
We shall then present  a somewhat different     argument
(based on   certain  plausible conjecture on the structure of the
beta-functions in the general bosonic or (1,1) 
supersymmetric sigma model)
which  will suggest   the 
 exactness of the NS-NS background \gme,\uga\
 already in the bosonic string theory.

\medskip 
Let us start with a   digression  on the 
structure of the type II superstring effective action.
By definition, the effective  action for the massless string modes 
is constructed so that to reproduce the string S-matrix. 
Field redefinitions \rf{\GW,\tsey,\mett,\metse}
 allow one to avoid ``quadratic'' or ``propagator correction''
  terms
(i.e. terms whose weak-field expansion  starts  with quadratic terms).
In addition, as is well known \GSW, 
 the  on-shell superstring  amplitudes  for massless modes 
do not contain 
(in contrast to the  bosonic string 
amplitudes) 
  $\a'$-corrections, 
i.e. the supergravity 3-point amplitudes are exact. 
This suggests  that there  may  exist a  field-redefinition choice 
(or, in the $\beta$-function context,   a 2-d RG  scheme)
   in which the weak-field expansion of the
$\a'$-dependent part of  the effective action starts 
with quartic  terms only
(i.e.,  $\a'^3 RRRR + ...$). 
Indeed, since  the  ``quadratic''
 and ``cubic''   terms 
that may be present in the action 
should not contribute to the S-matrix,  
  it may  be possible to  redefine them  away.
  More precisely, for our present purpose, 
  it is sufficient to argue  that  in  such   a scheme 
 the effective Lagrangian   will    not  contain terms like 
\eqn\laaa{
\Delta L = \   a_1  ( D...D R)^2  + 
a_2 ( D...D F )^2    
\  +  \  a_3 \  R_{\m .\n .} \ D...D F_{\m...}\   D...D F_{\n...} 
 \ .  } 
 Here  $R$ is  curvature, $D$ is covariant derivative and 
 $F = d C$ stands for any p-form field 
 strength of  the type II supergravity multiplet,
including $H_3$.
In this case there will be no terms like 
\eqn\cann{
   c_1 D...D R_{\m .\n .}
 +  c_2  D...D F_{\m...} D...D F_{\n... }  \  }
in the equation for the metric,  and  the terms like 
$ D...D F $ in the equation  for the p-form field.

To prove that the  terms in \laaa\  can  either be redefined away  
 or  modify the 
3-point  S-matrix
 (and thus are excluded in the superstring case), 
let us  make several observations.\foot{We  are grateful to 
 R. Metsaev  for an 
  important criticism  that helped to clarify the argument below.} 
We shall concentrate on the $a_3$-terms in \laaa\ since 
the argument
for $a_1-$ and $a_2-$ terms is trivial. 
First, as we will be interested only in the $c_2$-terms in \cann\  
which do not contain extra powers of $R$, 
we may  ignore commutators of covariant derivatives  in \laaa\
(it is, in fact, sufficient to replace $D_\m$ in \laaa\ by $\del_\m$). 
Second, the $a_3$-terms in \laaa\  which 
 contain (after integration by parts)  two contracted derivatives acting
on one  field, i.e. 
$ D^2 R$ or $D^2 F$ factors, can be eliminated by a  local 
metric or potential $C$ redefinition in the  standard 
$R + F^2$ kinetic term. Indeed,  
using Bianchi identities  and integrating  by parts, such terms  
can be put into the form proportional (up to higher-order
 terms that can be 
ignored)  to the leading parts of 
 equations of motion, i.e. they can be written as 
$R_{\m\n}  X_{\m\n}$  and $D_\m F_{\m\n...} Y_{\n...}$, 
and thus can be redefined away.  
Next, the only  $a_3$-terms in \laaa\ that may contribute 
to the on-shell 3-point amplitudes must have  all 
 indices of $D$-derivatives contracted 
  with field (polarization tensor)
 indices and not between themselves:
 for massless on-shell 
 amplitudes $p_1+ p_2 + p_3= 0, \ \ p^2_i =0, \ i=1,2,3$,  
  and thus $p_i \cdot p_j=0$.
  Let us consider, for concreteness,  the case of $F=F_3$.
  Then the $a_3$-terms that may contribute to the S-matrix can 
  only have at most two $D$-derivative factors
  (the total number  of indices of $h_{\m\n}$ and $C_{\m\n}$ in 
  the 3-point vertex 
  should be greater or equal to  the total number
  of derivatives). Explicitly, such terms are  
  \ $R_{\m\a\n\b} F_{\m\a\r} F_{\n\b\r}$
  and  $R_{\m\a\n\b} D_\r F_{\m\a\s} D_\s F_{\n\b\r}$
  (other contractions of derivatives are equivalent or 
  give vanishing contribution to the 3-point amplitudes). These    
   terms  can not be present  in the superstring effective action. 
   
  Finally, let us  show   that  all other possible 
  terms  with contracted  derivatives 
   that do not contribute to the S-matrix 
  can indeed be redefined  away. 
  Consider the generic case  of two contracted derivatives 
  (as already mentioned, 
  positions of derivatives in $D...D$ factors in \laaa\  are not
   important): \ 
  $ L_3= R_{\m .\n .}  D...D D_\l  F_{\m..}  D...D D_\l F_{\n..}$.
  Since  terms with $D^2 F_{\m..}$  can be redefined away, 
  we can write  $L_3=  R_{\m .\n .} D^2 ( D...D  F_{\m..} 
   D...D F_{\n..})$, 
  or, integrating by parts, as 
  $L_3= - D^2 R_{\m .\n .}  D...D  F_{\m..}  D...D F_{\n..}$.
  Since $D^2 R_{\m\b\n\r}  \to D_{\m} D_{\n} R_{\b\r} + ...$, 
     such terms can be again redefined away.

\medskip 

Returning to our main argument, let us now take into account 
 the specific form of the background in question  \gme\ and 
\spe.  
First, we  note that  its ``null'' structure  implies that 
all scalar invariants constructed out of 
 $R, D $ and    $F$  identically  vanish. 
The only non-zero components of $R$ and $F$ have, respectively, 
  two  and one of $u$-indices, and  the only non-zero component 
of the covariant derivative is $\del_i$ (the fields do not depend on $v$). 
This implies that the only non-trivial correction to the 
equation  $DF + ...=0$ for $F$   must be linear in $F$, i.e. it 
can only  have the structure $D...D F_{u...}$~. 
Such terms  that could
only originate
from the $a_2$ term in the effective 
action \laaa\ are prohibited by the above argument. 
Similarly, 
any  non-zero correction 
to the Einstein equation  must carry two $u$-indices.
Since the  curvature has only 
$R_{uiuj}$ components, the required 
2-nd rank tensor that can be constructed  from such curvature is 
$R_{uu}$ or $D...D R_{u.u.}$.
Thus corrections to the Einstein equation 
(or to the $\beta$-function for the metric) 
could only  be of the form
\eqn\couu{
d_1 D...D R_{u.u.}  +  d_2  D...D F_{u...} D...D F_{u...} \ .  } 
 They could only follow from the terms \cann\
in the covariant expression (note the position of indices on $F$-factors), 
which, however, 
should be  absent, as explained above,   in the  natural scheme 
in which the 2- and 3-point terms    in  
the effective action  are not modified from  their supergravity values. 
We conclude that in this  
 scheme the background \gme,\spe\ or \uga\ 
that solves the leading-order supergravity equations  of motion remains 
the solution to all  orders in $\a'$-expansion.
 
\medskip

Let us now present an alternative argument 
that will apply only to the NS-NS case  but 
will be valid also in the bosonic string  case: it 
 will  suggest  that the  NS-NS 
sigma model corresponding to 
\gme,\uga\ is conformal to all orders in $\a'$.
It is based  on  a  plausible conjecture \mett\
 that  for the  general  sigma model
$L= (G_{\m\n}+ B_{\m\n})(x) \del x^\m \del x^\n $
 there should exist  an  RG  scheme in which   all $\a'$
  corrections to the  beta-functions for $G_{\m\n}$ and $B_{\m\n}$ 
are at least {\it linear}   in the generalized curvature
 $\hat R_{+\m\n\l\r} $(=$\hat R_{-\l\r\m\n}$) 
     defined by  the connection 
$ \hat \Gamma^\l_{\m\n}  = \Gamma^\l_{\m\n} \pm  \ha H^\l_{\ \m\n}$.
This conjecture is supported,  in particular,  by   explicit
higher-loop computations
     \rf{\SSS}, and by the fact that parallelizable  spaces
     (corresponding to WZW models) are independently known to be
       finite \muk.
In such  a scheme all  corrections to the $\beta$-functions 
for $G_{\m\n}$ and $B_{\m\n}$ must be  of the
form
\eqn\corc{
\beta_{\m\n} = P_{\m\n}^{\k\l\r\s} (R, H,  D)\  \hat R_{+ \k\l\r\s} \ .
} 
In the present case of \gme,\uga\  one finds  that the only non-zero
components 
of the generalized curvature are (the Christoffel connection 
is $\Gamma^{i}_{uu} = - \ha \del_i K, \ 
\Gamma^{v}_{ui} =  \del_i K$ and $H_{ium}= \del_i b_m$)
\eqn\haa{
\hat R_{+ uiuj} = - \ha \del_i \del_j K  - \four   \del_i b_m \del_j b_m 
\ , \ \ \ \ \ \ 
\hat R_{+ unum} =   - \four  \del_i b_n \del_i b_m 
\ , \ \ \ \ \ \ 
\hat R_{+ jmiu} =  \del_i \del_j b_m \ .  } 
Note that $\hat R_{\m\n}=0$ on the equations of motion \asw.
Since the functions $K$ and $b_m$ do not 
 depend on $v$, and  any 
  corrections are possible only to   $\beta_{uu}$. 
For the present background  $D_u, D_m $ are 
 trivial  and $H_3$ and $\hat R$  have at least one 
$u$-index. This implies  that  possible corrections    
  must have the structure
$\beta_{uu} =  O_{ij} \hat R_{+uiuj}, 
$
where $O_{ij}$ is  a differential operator involving  $D_i$ only.
Since  here
$D_i = \del_i$, we get 
$O_{ij}= k_1\del_i \del_j +  k_2   \del_i \del_j \del^2  + ... $
(dots represent terms with higher  powers of $\del ^2$). 
As was noted    above, such ``propagator-correction'' terms 
can be in general redefined away
(they should, in fact, be absent in the minimal subtraction scheme). 


\bigskip

\subsec{\bf Conditions for space-time supersymmetry}

Let us now  determine the conditions under which 
the  background \gme,\uga,\asw\ 
preserves a fraction of type II space-time supersymmetry. 
Explicit examples will be given later in section 3.

In the case of the metric and $H_3$ given by  \gme,\uga\ and constant 
dilaton
the type IIB  dilatino transformation law gives   the condition
\eqn\dili{
\del_i b_m  \Gamma^u \Gamma^{im}\ep =0\ .}
The  condition of the 
 vanishing of the  gravitino variation is\foot{Below the sign
 of $H_3$-term is chosen as plus; the case of the minus sign
 (corresponding to $b_m \to -b_m$)  leads to  equivalent
 conclusions.}
  \eqn\lop{
  [\del_\m +
  {1 \ov 4}  ( \omega_{\hat \l\hat\r \m} \pm   \ha  H_{\hat \l\hat\r\m }
)
  \G^{\hat \l\hat\r} ] \ep=0 \ . }
 The  $v$-component  is  solved
provided
$\del_v\ep =0$, i.e. $\ep=\ep(u,x,y)$.
The $i$ and $m$ components  give\foot{
 We use that  $H_{ium} = \del_i b_m$ and 
  the  tangent-space components
of the Lorentz connection given in eq. (4.5) below.
We also  omit  hats on indices of $\Gamma$-matrices as the 
corresponding vierbein components are trivial.
} 
\eqn\gravi{
\del_i \ep +  {1\over 4}  \del_i b_m \Gamma^{um} \ep=0\  ,
}
\eqn\gravix{
\del_m \ep -  {1\over 4}  \del_i b_m \Gamma^{ui} \ep=0\  .
}
{}For non-constant $H_3$, i.e.
 $\del_i \del_j b_m\not=0$, 
 these equations imply\foot{This is the integrability condition
for \gravi,\gravix\ which  is found 
by multiplying \gravi\ by $\del_m$,  \gravix\  by 
$\del_j$,  subtracting and using \gravi,\gravix\  again.} 
\eqn\nuevas{
\Gamma^u \ep =0\ .
}
As a result, $\ep $ is independent of $x^i$ and $y^m$.
In general, the  $u$-component of the gravitino variation gives
\eqn\gravu{
\del_u \ep + {1\over 4}\del_i K\Gamma^{ui} \ep
 -  {1\over 4} \del_i b_m \Gamma^{im} \ep=0\ .\ \
}
In view of  eq.~\nuevas\  and the fact that $\ep $ is independent of
$x$ while $\del_i b_m$ is a function of $x$, we get the condition
\eqn\qwe{
\del_i b_m  \Gamma^{im} \ep  = 0\ , 
} 
as well as  $\del_u \ep =0$. 
 
The  condition \nuevas\ (or  \qwe ) 
 ensures   the  vanishing of the dilatino variation \dili.
The number of  remaining supersymmetries 
thus   depends  on  existence of constant $\ep$  solutions 
of \qwe.

One important  conclusion is that  the pp-wave background \gme,\uga\ 
with non-constant $H_3$ 
(or its R-R counterpart with $H_3 \to F_3$ discussed below) 
does not have ``supernumerary''  
supersymmetries for which  $\G^u \eps\not=0$
(i.e. solutions for $\ep$ 
which are ``orthogonal'' to the ones 
satisfying the  GS \lc  gauge condition \nuevas).   
As a result, in contrast to the case considered in \mam, 
we should not expect to find extra linearly realized 
supersymmetries in the 
corresponding \lc GS action  \refs{\mald,\cvet,\mam,\hsu}. 
Still, the  \lc actions we shall get will  have 
several  features in common with the one corresponding to the 
$F_5$ case  discussed in \mam.

\medskip

The above discussion applied 
to the case of the $H_3$ background 
of the form given in \uga.
In general, one  can  show  that  there are no 
 ``supernumerary'' supersymmetries
for any pp-wave  background  
supported by any  {\it non-constant}  
 $H_3$ form.
Indeed, consider  the most 
general ansatz for the ``null' 2-form field:
$B_2= b_s(x)\ du \wedge dx^s$.
Here  $x^s$ are all 8 transverse coordinates, i.e. this 
 includes  \uga\ as a special case.
Then $H_{usr}\equiv f_{sr}=\del_s b_r -\del_r b_s$, and thus 
 the $s$-component of the gravitino equation gives
(cf. \gravi,\gravix) 
$$
\del_s \eps +\four f_{sr}(x) \G^{ur} \eps = 0  \ .
$$
Let us assume  that, e.g.,  $f_{12}\not=0$.
Acting on  the $s=1$ equation  by $\del_2$ and on the
$s=2$ equation  by $\del_1$, and subtracting, we get
$\del_s f_{12} \G^{us} \eps = 0$.
{}For non-constant $f_{12}$, this can be satisfied only if 
 $\G^u \eps =0$.
Again, this 
implies \refs{\mald,\cvet,\mam}  that the \lc GS Lagrangians
corresponding  pp-wave backgrounds
of the form $ds^2=dudv+K(x)du^2+dx_s^2$ supported by
non-constant  $H_{usr}$ or $F_{usr}$ forms
will not have  
linearly-realized  world-sheet supersymmetry.  
  
The case \lii\ of constant $H_3$  (i.e. $\del_i \del_j b_m =0$)
is special.
 Here the gravitino condition 
does not reduce the number of supersymmetries.
  For generic  constant $H_3$ configuration in type IIB theory 
  the dilatino variation equation reduces
the number of unbroken supersymmetries to 16, as one is
to impose     $\Gamma^u \epsilon =0$     to satisfy \dili.
However, for  special  ``self-dual''   matrices $\del_i b_m$
(which correspond, in particular, to
 the case of  the  direct sum of the two  Nappi-Witten \nw\ 
 models which is the Penrose limit of the $AdS_3\times S_3$
background)
 the condition \dili\  breaks less than 16 supersymmetries.\foot{  
 In the example of  the Penrose limit of $AdS_3\times S_3$,  the
resulting number
 of unbroken supersymmetries, i.e.  the solutions to 
$\Gamma^u (1-\G_{1234})\ep=0$,  is 24.}

\newsec{pp-wave backgrounds  parametrized by holomorphic functions}

Here we shall specialize to a particular subset of the above backgrounds
which 
may be viewed as  NS-NS analogs of the R-R 5-form backgrounds
discussed in \mam. 
They can be found from  the  general exact solution \gme--\gensl\ 
with  the even  number $d=2n$  of $x_i$ coordinates
and thus   $8-2n$ ``spectator'' coordinates $y_m$ 
organized  into two sets of $n$  and $4-n$ complex coordinates 
$z_a$ and $z_\a$, respectively. 
The Laplace equation
for $b_m$ in \asw\  can then be solved in terms of holomorphic and anti-holomorphic
functions.
The general structure of the solution  written in complex  coordinates 
is then
\eqn\metra{
ds^2=dudv + K(z_a, z^*_a)du^2+dz_adz^*_a+dz_\a dz_\a^*\ ,
}
\eqn\gauga{B_2 = \ha \eta_\a (z_b)\  du\wedge   dz_\a
 +c.c.\ ,     \ \ \ \ \ \
H_3= - \ha  du\wedge \omega_2\ ,\ \ \ \ \ 
\omega_2=\del_a\eta_\a (z_b)\ dz_a\wedge dz_\a+c.c.\ ,
}
$$ \eta_\a =\eta_\a (z_1,...,z_n) \ ,\ \ \ \   \ \ \   a=1,...,n\ ,
\ \  \a=n+1,...,4\  . $$
The cases of  $n=1,2,3$ represent inequivalent solutions.
A more general solution 
(belonging again to the family of solutions of section 2.1) 
can be obtained by choosing\foot{The particular case $\eta_\a=0$ is equivalent to the solution 
\gauga\ by the simple change of coordinate $z_\a\to z_\a^*$.}
\eqn\oopp{
\omega_2=\del_a\eta_\a (z_b)\ 
dz_a\wedge dz_\a+\del_a\rho_\a (z_b)\ dz_a\wedge dz^*_\a+c.c.
\ . } 

\subsec{\bf Particular examples  }

Simple examples of  solutions  are  found by choosing:
\eqn\eneuno{
n=1\ :\ \ \ \ \ \omega_2=\del_1 \eta (z_1)\ dz_1\wedge dz_3+c.c.
}
\eqn\enedos{
n=2\ :\ \ \ \ \ \omega_2= \del_1 \eta (z_1)\ dz_1\wedge dz_3+
\del_2 \tilde \eta (z_2)\  dz_2\wedge dz_4
+c.c.
}
\eqn\enetre{
n=3\ :\ \ \ \ \ \omega_2= \del_1 \eta (z_1,z_2)\ dz_1\wedge dz_4+
\del_2 \eta (z_1,z_2)\ dz_2\wedge dz_4 +
c.c.
}
Eq.  \eneuno \ is a special  case of \gauga\ with $\eta_3=\eta(z_1)$
and $\eta_2=\eta_4=0$. The $n=2$ example  \enedos\ is 
obtained  from \gauga\   by setting $\eta_3=\eta(z_1)$ and
$\eta_4=\td \eta (z_2)$.
The more general $n=3$ case has $\eta=\eta_4 (z_1,z_2,z_3)$ (then
  $\omega_2$ has one
extra term containing $dz_3$).

{}For the examples in  \eneuno , \enedos , eq.~\asw\ has 
the following special solutions for $K$ (we choose $K_0=0$ in  \gensl )
\eqn\sola{
n=1\ :\ \ \ \ \ \ \ K=-{1\over 2}|\eta(z_1)|^2
\ ,
}
\eqn\solat{
n=2\ :\ \ \ \ \ \ \ K
=-{1\over 2} (  |\eta(z_1)|^2 + |\td \eta(z_2)|^2    ) \ .
}
It is useful also to record  the form of  $H_3$ 
in \eneuno\ and \enedos\  in real coordinates.
Defining 
$$
z_1=x_1+i x_3\ , \ \ \ \ z_2=x_5+ix_7\ ,\ \ \ \  z_3= x_2+ix_4\ ,\ \ \ 
\ z_4=x_6+i x_8\ ,
$$
we get
\eqn\caruno{
n=1 :\ \ \ \omega_2=(\del_1 \eta +\del_1^* \eta ^*)  
( dx_1\wedge dx_2 - dx_3\wedge dx_4) + i(\del_1 \eta -\del_1^* \eta ^* )
( dx_1\wedge dx_4 + dx_3\wedge dx_2) 
}
and
$$
n=2 :  \ \ \ \omega_2=(\del_1 \eta +\del_1^* \eta ^*)  
( dx_1\wedge dx_2 - dx_3\wedge dx_4) + i(\del_1 \eta -\del_1^* \eta ^* )
( dx_1\wedge dx_4 + dx_3\wedge dx_2) 
$$
\eqn\cardos{
+ \ (\del_2 \td \eta +\del_2^* \td \eta ^*)
( dx_5\wedge dx_6 -dx_7\wedge dx_8) + i(\del_2 \td \eta -\del_2^* \td
\eta ^* )  
( dx_5\wedge dx_8 + dx_7\wedge dx_6) 
}
Here for convenience we are using 
the same notation $x_i$ for all of the transverse coordinates, instead of the splitting
them into $ (x_i, y_m)$ as we did  in section 2. 

It is easy to read off the relation between $\eta_\a $ in \gauga\
\ and $b_m $ appearing in the general expression \uga .
{}For example, in the $n=1$ case \caruno ,
 $x_2,x_4$ play the role of the two ``free'' 
$y_m$ coordinates and
we have two components of $b_m$
which depend on ``dynamical'' coordinates $x_1,x_3$ 
 $$ b_2 = b_2(x_1,x_3)=
{\rm Re}\ \eta\ ,\ \ \ \  \ \ \  b_4=b_4(x_1,x_3)= 
- {\rm Im}\ \eta \ , \ \ \ $$
so that 
\eqn\rekl{  
\del_1 b_2 =-\del_3 b_4 ={\rm Re}\ (\del_{1} \eta)= \ha 
(\del_{1} \eta +\del_1^* \eta ^*) \ ,\ \ \ \ 
\del_3 b_2 =\del_1 b_4  =- {\rm Im}\ (\del_{1} \eta)= 
     \ha i  
(\del_{1} \eta -\del_1^* \eta ^*)
\ . }

\bigskip
\subsec{\bf Space-time supersymmetry}

Let us now count the number of unbroken supersymmetries for these
solutions, 
solving the conditions \nuevas,\qwe\  in the above special  cases.
First, the condition \nuevas\ breaks 16 
supersymmetries.
Consider now the remaining equation \qwe .
In the $n=1$ case, using \caruno , we get the restriction\foot{
{}From  \caruno\ 
we find that \qwe\ takes the form  
$ [a(x) (\G_{12} - \G_{34})  + b(x)  (\G_{14} - \G_{23})] \ep=0 $  
which can be satisfied for any functions $a,b$ if we require 
$  (\G_{12} - \G_{34}) \ep=0 $ and   $(\G_{14} - \G_{23}) \ep=0$.
Each of these two  conditions is  equivalent to $(1+ \G_{1234}) \ep=0
$.}
\eqn\soly{
(1+ \Gamma_{1234})\ep =0
}
which breaks 8 more supersymmetries, so that
 there are 8 remaining   supersymmetries.
In the $n=2$ case, using \cardos , we obtain 
\eqn\solq{
(1+ \Gamma_{1234})\ep =0\ ,\ \ \ \ (1+ \Gamma_{5678})\ep=0\ , \
}
and, as  a result, there are  4 unbroken supersymmetries.

In a generic $n=1$ model, one has
\eqn\genn{
\omega_2= \del_1\eta _\a(z_1) dz_1\wedge dz_\a +c.c.  \ ,
\ }
where  $\eta_\a $ ($\a=2,3,4$) are  3 independent  functions.
Here  the condition 
\qwe \  becomes 
\eqn\solyy{
(1+ \Gamma_{1234})\ep =0\ ,\ \ \ \ (1+ \Gamma_{1537})\ep =0\ ,
\ \ \ \ \ (1+ \Gamma_{1638})\ep =0\ .\
}
This leads to  two unbroken supersymmetries.
The same conclusion is reached for  a generic $n=3$ model  with
$
\omega_2= \del_a\eta_4 (z_1,z_2,z_3) dz_a\wedge dz_4 +c.c.
$, where   \qwe \ leads again to three conditions of the form
\solyy , i.e. to   two unbroken supercharges.

The generic $n=2$ case
$$
\omega_2= \del_1\eta_3 (z_1,z_2) dz_1\wedge dz_3 
+\del_2\eta_3 (z_1,z_2) dz_2\wedge dz_3 
$$
\eqn\wri{
+\  \del_1\eta_4 (z_1,z_2) dz_1\wedge dz_4 +
\del_2\eta_4 (z_1,z_2) dz_2\wedge dz_4 + 
c.c. }
requires a closer examination.
By writing \wri\  in Cartesian coordinates, we find  that
for arbitrary functions $\eta_3 (z_1,z_2), \eta_4 (z_1,z_2)$,
the
supersymmetry condition \qwe\  gives
\eqn\solyz{
(1+ \Gamma_{1234})\ep =0\ ,\ \ \ \ (1+ \Gamma_{5274})\ep =0\ ,
\ \ \ \ (1+ \Gamma_{5678})\ep =0\ ,
\ \ \ \ \ (1+ \Gamma_{1638})\ep =0\ .\
}
Since the  last condition  follows from the first three,
we conclude that the generic $n=2$ case
also preserves two supersymmetries.

\newsec{String sigma model actions for the NS-NS  pp-wave
backgrounds}

\subsec{\bf Covariant action }

The bosonic part of the
sigma model Lagrangian corresponding to the generic background \gme ,
\uga , \gensl\
(with $K_0=0$)
is given by \foot{ We
shall use Minkowski world-sheet coordinates with $\s^\pm =\tau\pm
\s $, and $\del_\pm = \ha (\del_\tau  \pm \del_\sigma $). The string
action
is $S= {1 \ov \pi \a' } \int d^2 \s \ L$. The space-time \lc
coordinates are $u=y-t, \ v= y +t $. } 
 $$ L_B = \del_+ u \bd
v - \ha b^2_m(x)\del_+u \bd u +b_m(x) (\del_+ u \bd y_m -
 \del_+ y_m \bd u ) $$
 \eqn\stre{
 + \  \del_+ x_i \bd x_i + \del_+ y_m \bd y_m\ . 
}
Note that by applying 2-d duality (i.e.  T-duality in the target space)
in $y_m$ 
we get a model with zero 3-form field but off-diagonal metric, 
i.e. \stre\ becomes
$$ \td L_B = \del_+ u \bd
v - \ha b^2_m(x)\del_+u \bd u +b_m(x) (\del_+ u \bd \td y_m + 
 \del_+ \td y_m \bd u ) $$
 \eqn\stre{
 + \  \del_+ x_i \bd x_i + \del_+ \td y_m \bd \td y_m\ . 
}
In general, the fermionic
  part
 of the
 (1,1) world-sheet supersymmetric sigma model can be   written in terms
of the
 generalized Lorentz connections
 $  \omega ^{\hat\rho }_{\pm \hat \nu\mu }=
\omega ^{\hat \rho}_{ \hat \nu\mu }\pm \ha H^{\hat \rho}_{\ \hat \nu\mu
}$\ 
\eqn\ferm{
L_{F}=
i\lambda_{R\hat\rho }(\delta^{\hat \rho }_{\hat \nu}
\del_+ +  \omega ^{\hat \rho}_{- \hat \nu\mu }
\del_+ x^\mu )
\lambda_{R}^{\hat \nu} + 
i\lambda_{L\hat \rho  }(
\delta ^{\hat \rho }_{\hat \nu}
\del_- 
+ 
\omega ^{\hat \rho}_{+ \hat \nu\mu }
\del_- x^\mu )
\lambda_{R}^{\hat \nu} + \ha \hat R _{+ \hat \rho \hat \nu \hat \sigma
\hat \m}
\lambda_{L}^{ \hat \rho} \lambda_{L}^{\hat \nu}\lambda_{R}^{\hat \sigma
}
\lambda_{R}^{\hat \m}  \ .
}
 In the present case of  \gme ,\uga ,\gensl\ \foot{Here 
  $\eta_{\hat u \hat v} = \ha, \  \eta^{\hat u \hat v} = 2,$ etc,
and hats on $u,i,m$ indices can be omitted.}
\eqn\jor{
 \omega _{\pm \hat u\hat m}= \pm \ha \del_i b_m dx_i
 \ ,\ \ \ \ \  \omega _{\pm \hat i\hat m}=\mp \ha \del_i b_m \ du \
  ,\ \ \ \ \
 \omega _{\pm \hat u\hat i}= \ha \del_i K du \mp \ha \del_i b_m dy_m \ , 
}
and the non-zero components of $\hat R_{+uiuj}=  \hat R_{-uiuj}$
were given in \haa.


The explicit form of the Lagrangian 
\stre\ in the  case of the 
$n=1$ solution \metra,\gauga,\eneuno \  parametrized by an 
arbitrary 
holomorphic function $\eta=\eta(z_1) $ is 
($\a=2,3,4$) 
$$L_B = \del_+ u \bd
v - \ha \eta \eta^*\del_+u \bd u + \ha \eta (\del_+ u \bd z_3 - 
 \del_+ z_3 \bd u )+ \ha \eta^* (\del_+ u \bd z_3^*  - 
 \del_+ z_3^* \bd u )
$$  \eqn\shhh{ 
+\ \del_+ z_1 \bd z_1^* + \del_+ z_\a \bd z_\a^* \ . }
Similarly, 
by applying T-duality transformations
in 
$z_3=x_2+ix_4$  we get a  special case of pure-metric model \stre\ 
$$ 
\td L_B = \del_+ u \bd
v - \ha \eta \eta^*\del_+u \bd u + \ha \eta (\del_+ u \bd z_3 + 
 \del_+ z_3 \bd u )+ \ha \eta ^* (\del_+ u \bd z_3^*  + 
 \del_+ z_3^* \bd u )
$$\eqn\stretd{  
+\ \del_+ z_1 \bd z_1^* + \del_+ z_\a \bd z_\a^* \ . 
}

\subsec{\bf Light-cone gauge action}
Let us now consider the form of the string Lagrangian
in   the light-cone gauge. Since $u$ and $ \lambda_{L,R}^u$
obey free field equations, we can supplement the superconformal gauge
 with  the standard light-cone gauge conditions
\eqn\lcgg{
u=2\a' p^u \tau\    , \ \ \ \ \ \ \  \lambda_{L,R}^u=0   \ .
}
Then the  bosonic part of the
Lagrangian \stre\   takes the form
\eqn\sbbe{ L_B =  \del_+ x_i \bd x_i 
- \ha \mm^2 b^2 _m(x) - \mm b_m(x) (\del_+ y_m - \del_-  y_m )
+ \del_+ y_m \bd y_m
\ ,}\eqn\maas{
 \mm\equiv \a' p^u = \del_\pm   u \ .}
Note that in our notation $u,v,x^i,\sqrt{\a'},(p^u)^{-1}$ 
and thus $\mm$ have 
dimension of length, while the world-sheet coordinates 
$\tau$ and $\sigma$$\in [0,2\pi)$ are dimensionless.\foot{The standard 
alternative is to rescale $\tau$ and $\sigma$ by $\a' p^u$ 
(which is a symmetry of the $2+d$ dimensional conformal theory), 
thus giving them dimension of length. Then  in the \lc gauge 
 $x^+ = 2 \tau, \ \
 0\leq \s < 2 \pi \a' p^u$.} 
The components of the metric and 2-form tensor, i.e. 
$K$  and  $b_m$ in \gme,\uga\ or $\eta_\a (z)$ in \gauga, 
are also dimensionless.
{}From the world-sheet point of view it is more natural to  treat 
$u,v,x^i$ as dimensionless while $\tau,\s$ as having dimension of length.
Then $\mm$ has world-sheet dimension of  mass.

In view of \asw,\haa, the fermionic part of the action \ferm\
becomes 
quadratic in fermions 
    \eqn\sffe{ L_F = i \big[ 
      \l_R^i \del_+   \l_R^i    
        +     \l_L^i \del_-   \l_L^i
   +  \mm \del_i b_m(x)  \ (  \l_R^i    \l_R^m - \l_L^i    \l_L^m )
  +     \l_R^m \del_+   \l_R^m    +     \l_L^m \del_-   \l_L^m \big]  \ . }
The sigma model action \sbbe,\sffe\ follows 
also directly from the (1,1) 
superfield form of the  action 
$\int d^2 \s d^2\vartheta (G_{\m\n} + B_{\m\n})(\hat X)  D_+ \hat X^\m D_-
\hat X^\nu $
with $B_{um} = b_m$ and the superfield $\hat X^u$ chosen in the \lc gauge 
form 
$\hat X^u=u =  2 m \tau$. The latter choice 
breaks the 2-d Lorentz
invariance and 
the  (1,1) 2-d supersymmetry. 

The resulting ``transverse'' gauge-fixed 1+1 dimensional theory   is
thus not
2-d Lorentz-covariant -- 
  the bosonic term originating from the $B_2$-coupling contains
explicit 
sigma-derivative (or $\del_+ -\del_-$) term.
The  absence of manifest 2-d Lorentz  covariance of the ``transverse''
theory
is not unfamiliar:
it is  generic to many  similar pp-wave string models  written in
the light-cone gauge.
In particular, the  Lorentz  covariance is absent in all cases where $K$
has $u$- (and thus $\tau$-) dependence.

The Lagrangian  \sbbe,\sffe\ may be interpreted as describing a 
 system 
of chiral scalars $y^m_{L,R}$ and their (1,0) and (0,1) superpartners 
$\lambda^m_{L,R}$ interacting  with scalars $x^i$ and 
fermions $\lambda^i_{L,R}$.
There is an obvious  left-right decomposition in the $y^m,\l^m$ sector,
but the two chiral sectors are mixed by 
the interaction terms  in the $x^i, \l^i$ sector.

\newsec{Light-cone gauge GS string actions 
for the non-constant   R-R pp-wave  backgrounds
}
By applying  S-duality transformation to the background 
 \gme , \uga, \gensl , i.e. replacing $H_3$ by $F_3$, 
we obtain  another  solution of type IIB supergravity 
theory which has  constant dilaton and
non-trivial R-R 3-form  field.
According to the argument in section  2.2, this R-R background, 
just like its NS-NS counterpart, 
 is, in fact,  
an exact solution of type IIB string theory. 
The conditions of space-time supersymmetry are again determined by
\nuevas\ and \qwe . 

To find the form of the corresponding \lc gauge Green-Schwarz action 
we follow the same logic as was used in \refs{\mt,\RT}.
As was mentioned in \refs{\mt,\RT} in the case of 
 constant R-R  p-form  strengths, and as we shall explicitly 
prove below in the general case of non-constant R-R pp-wave backgrounds, 
 the property that the curvature and $F_p$ have 
null structure implies that  all higher than quadratic 
 terms in fermions 
should be   absent from the GS action written in the light-cone gauge
\eqn\rrty{\Gamma^u \theta^{1,2} 
= 0\ , \ \ \ \ \ \ \ \ \ \
u=2\mm \tau\ ,\ \ \ \ \  
\mm\equiv \a' p^u \ ,  } 
while the quadratic fermionic term is essentially determined \refs{\mt,\RT}
(see also \refs{\MMM,\cvst}) 
 by 
the structure of the corresponding generalized covariant derivative\foot{Here 
we use the standard normalization in which $e^\phi=1$, 
so that the R-R terms in the covariant derivative in eq. (5.4) 
of \RT\ should be multipled by extra 1/2.}
\eqn\quaf{
L_{2F} =i \del_- X^\m\ \bar \theta^1 \G_\mu \hat D_+ \theta^1\ +\ 
i \del_+ X^\m\ \bar \theta^2 \G_\mu \hat D_- \theta^2 \ ,  }
where in  the $F_3$-case
\eqn\genk{
\hat D_\mu \theta^{1,2} =
(\del_\m + \four \omega_{ \s \l  \m} \G^{ \s  \l } )\theta^{1,2} 
-  { 1\ov 8 \cdot 3!} F_{\s\n\l} \G^{\s\n\l}\G_\m \theta^{2,1} \ .   }

\subsec{\bf Light-cone gauge GS action is quadratic 
in fermions for generic  pp-wave background}
The form of the covariant GS action in a generic type II supergravity 
background  is very complicated, containing terms 
of all possible  powers in  the two 10-d MW spinor variables $\theta^I$ 
which come out of the component expansions of the superfields
entering the superspace  form of the action \howe. 
For certain   backgrounds, 
the action written  in  a special  $\k$-symmetry gauge 
may become quartic in fermions (as is the case 
 for the  $AdS_5 \times S^5$ action in the \lc gauge \mtt, see also \hhh). 
A particularly simple case is that of the 
homogeneous plane-wave  backgrounds \blau\ 
-- here the \lc gauge  GS action turns out to be 
 quadratic in fermions  \refs{\rrm}.

Below we shall prove  that this is true 
also for generic inhomogeneous ($R\not=\const, \ F_p\not=\const$) 
pp-wave backgrounds  with the metric admitting a covariantly constant null Killing vector (i.e., for example, \gme) and the R-R or NS-NS 
p-form strengths having the ``null'' structure \spe. 
The key property we will use is that the curvature and the p-form
strengths do not have lower $v$-components and do not depend on $v$. 
Another important  point  is that the curved-space 
GS action (which is  a ``supersymmetrization''
 of the standard bosonic sigma model action) is quadratic 
in 2-d derivatives $\del_a$. This implies that  generic  
 fermionic terms  can only be  of the following 3 types 
$$L_1= D^n R ... D^k F ...  \bar \t... \t ...\bar  \t ...\t  \  
 \bar  \t ... \del \t \ \del X   \ ,  \ \ \ \ \ \ \
 L_2= D^n R ... D^k F ... \bar  \t... \t ...\bar \t ...\t \
  \del X \del X\
   \ , 
 $$
\eqn\suiy{   L_3= 
D^n R ... D^k F ... 
\bar \t ... \t ... \bar \t...\t\   \bar \t ...\del  \t\    \bar 
 \t ... \del \t    \ , } 
where $X$ are  the bosonic coordinates $X^\m= (u,v, x^i)$, 
and dots between $\bar \t$ and $\t$ 
stand for products of  $\G$-matrices. 

To get a non-zero result
from a particular term 
after imposing the \lc gauge condition 
 $\G^u \t=0$  each 
$\bar \t ... \t$ factor in \suiy\ must have the form 
 $\bar\t  \G^v \G^{i_1} ...\G^{i_n} \t$.
 Since the background-dependent factor $D^n R ... D^k F $ cannot have 
lower $v$-indices, each $\G^v$ must be accompanied by  
 $\del X^u\equiv \del u$. That immediately  implies 
that all $L_3$-terms in \suiy\  must vanish, while 
$L_1$ must be quadratic 
and $L_2$ --  at most quartic in $\t$. 
 In addition,  all non-trivial 
$D^n R ... D^k F $ factors 
 must contain at least one lower $u$-index which is to be contracted
 with either $\del X^u$ or  some $\G^u$ in the fermionic factor;
in the latter case, such term vanishes.\foot{More precisely, 
$\del X^u$  may be also  contracted  with $\G_u$  between 
$\bar \t$ and $\t$ 
  but, since $\G^u \G_u=2$  when acting on fermions subject to 
the \lc gauge condition,
this  leads to an  equivalent conclusion.}

 This  implies the 
vanishing of  all ``non-flat''  terms in $L_1$ and 
 quartic fermionic terms in $L_2$, 
and thus 
leaves us  with   the standard  flat-space 
GS term 
$L_1 = \del u \bar \t \G^v \del \t$ 
as well as with   the following candidates
 for the non-vanishing  quadratic fermionic terms  
\eqn\cand{
L_2=  D_{i_1} ... D_{i_m}  F_{u j_1...j_{p-1}}  \ 
  \del u  \   \bar  \t   \G^v  \G^{i_1}  ...\G^{j_{p-1}}  \t \ . }  
Furthermore, the 
 only term of that type that can actually appear 
in the GS action should contain {\it no} 
 covariant derivatives acting 
on $F_p$ --
each covariant derivative would be accompanied by 
an extra $\bar\t ...\t$ factor. 
Indeed, this follows simply from dimensional considerations:
if $X$ has dimension of length $l$ (e.g., $\sqrt {\a'})$, then  
$\t$ has dimension $l^{1/2}$, the metric and $(p-1)$-form potentials are dimensionless, and thus   $D_\m$  and $F_{\m_1...\m_{p}}$  have dimensions 
$l^{-1}$ (while  $R_{\m\n\l\r}$ -- dimension  $l^{-2}$).

As a result, we arrive at   the same form of the action 
as in the case of the constant null  R-R  flux  \refs{\rrm,\mt,\RT}: 
 the only   non-trivial  coupling to the background field  strength 
is through the generalized covariant derivative that enters the
 gravitino supersymmetry transformation rule, i.e. 
\genk\ (with other p-form terms  and the dilaton $e^\p$ factors 
included in general \john).

To summarize, the  above argument   allows one to  determine  the form of the 
\lc gauge GS action  for any inhomogeneous  R-R pp-wave 
background, in particular, for  the $F_5$-form background in \mam. 
The general structure of the GS action corresponding to the metric 
\gme\ supported by a R-R background $F_p (x)$ of the form  \spe ,
which solves the 
supergravity  equations of motion (i.e. $R_{uu} \sim F_{u...} F_{u...}, 
\ \  \del_i F^{iu...} =0 $), written 
in the \lc gauge \rrty\ 
is thus  (ignoring explicit value of the  numerical 
normalization factor $c_p$ in the last term)
\eqn\syym{
L= \del_+  x_i  \del_- x_i    +  \mm^2 K(x) 
+ i \theta^1  \G^v \del_- \theta^1  + i \theta^2 \G^v  \del_+ \theta^2  
+ i \mm  c_p F_{ui_1...i_{p-1}}(x) 
\ \theta^1 \G^v  \G^{i_1...i_{p-1}}\theta^2 \ . }

\subsec{\bf UV finiteness  of the light-cone  GS theory}

Viewing \syym\  as a 2-d field theory, it  is natural to assign 2-d 
length dimensions to $\tau$ and $\s$  and thus to assume that 
other dimensions are 
 $[x]=0, \ [\theta] = - 1/2 , \ [\mm] = -1 $.
Then $K(x)$ and $F_p(x)$ are dimensionless, and dimensional analysis 
implies that  the only possible logarithmically\foot{Trivial quadratic divergences cancel 
because  of equal total numbers of bosons and fermions.}
divergent $l$-loop 
counterterms in  this theory
should  be proportional to $\mm^2$.
Therefore, they  must  be linear in derivatives of  
$K$ and quadratic in 
derivatives of $F_p$, 
i.e.  
\eqn\dive{
\Delta L^{(l)} = \mm^2 \big[ a_l\  \del^{2l}  K  (x)
 +  b_l \ \del^{l-1} F_{ui_1...i_{p-1}}(x) \  
\del^{l-1} F_{ui_1...i_{p-1}}(x)     \big]  \ ,  \ \ \ \ 
\  \ \ \ l=1,2,... \ ,  } 
where $\del^{l-1} $ stands for $\del_{j_1} ... \del_{j_{l-1}} $, etc. 
These are the same counterterms as expected \laaa,\couu\ on the  general grounds in the covariant theory before the \lc gauge fixing  
(note that here $R_{uiuj} = -\ha \del_i \del_j K$ and $\mm = \del_\pm  u$). 
The coefficients in \syym\ are such that the 1-loop
divergences cancel (due to the relation between $K$ and $F_p$
as in \asw,\gensl).
Denoting by $G(\xi-\xi')$ the   propagator  of
 the 2-d bosons $x_i(\xi)$ \ ($\xi^a= (\tau,\s)$), 
the propagators of the fermions $\theta_L$ and $\theta_R$ 
are then $\del_+ G(\xi-\xi')$  and $\del_- G(\xi-\xi')$, 
and thus the coefficients in \dive \  are 
$$ a_l \sim \int d^2\xi\ [G(\xi -\xi')]^{l}_{\xi\to \xi'} \ , \ \ \ \  
 b_l \sim \int d^2\xi  d^2\xi'\  [G(\xi -\xi')]^{l-1}
\del_+ G(\xi-\xi') \del_- G(\xi-\xi')\ . $$
Integrating by  parts in the expression for $b_l$ we can transform it 
to the same form as $a_l$:
\  $  b_l \sim \int d^2\xi  d^2\xi'\  [G(\xi -\xi')]^{l } 
\del_+ \del_- G(\xi-\xi') \sim \int d^2\xi\   [G(\xi -\xi')]^{l}_{\xi\to \xi'} $. 
Since $\del_{i_1} F_{ui_{1}...i_{p-1}} =0, \ \del^2 F_{ui_{1}...i_{p-1}}=0
$ the  cancellation seen at 1-loop order 
should continue at  higher loops.\foot{In a particular  (e.g. dimensional) 
regularization scheme  
all higher tadpoles $a_l, \ l >2,$ can be set equal to zero 
(they lead only to higher than  first powers 
of logarithm  of the UV cutoff and
 thus can be ignored). 
 In the
case when the theory has extended  2-d  supersymmetry the 
cancellation of the two 
contributions in \dive\ can be understood as a consequence 
of non-renormalization of the chiral superpotential. 
This is what happens, e.g.,  
in the (2,2) supersymmetric model of  \mam\  where the bosonic potential 
is 
$  |\del_i W|^2$ and the Yukawa coupling matrix is $\del_i \del_j W$.}

This conclusion is of course  in agreement 
with the general  finiteness argument 
given in section  2.2. 
As a result, the \lc gauge theory is UV finite 
(but its  scale invariance explicitly broken by the 
presence of the  ``mass''  parameter $\mm$).

\subsec{\bf Explicit form of the \lc  GS action 
for the  R-R  3-form pp-wave background}
Let us now  return to the specific  case of our interest, 
namely, the background \gme,\uga\ with $H_3$ replaced by $F_3$. 
 As in \RT, 
we shall  keep  the free-theory notation 
 $\theta^1\equiv \theta _L$, $\theta^2\equiv  \theta _R$.
  Then the  corresponding light-cone  gauge GS Lagrangian is given by the
sum of the following bosonic and fermionic parts
(cf. \sbbe,\sffe,\quaf)\foot{We follow the spinor notation of \RT, 
i.e. switch to the 16-component notation 
for the spinors, with $\gamma^\m$  being  16-component real and symmetric matrices 
which replace 32-component matrices $\G^\m$.
}
\eqn\strr{ L_B = \del_+ x_i \bd x_i - \ha \mm ^2 b^2_m(x) 
 + \del_+ y_m \bd y_m\ , 
}
\eqn\ferr{
L_{F}=  i \theta _R \gamma^v \del_+ \theta_R + i \theta_L \gamma^v \del_-
\theta_L 
- 
\four  
i  \mm \del_j b_m(x) \  \theta_L \gamma^v \gamma^{jm}\theta_R\ . 
}
As usual, we rescaled  the fermions by  power of $p^u$.
For comparison,  the  fermionic term in the \lc   GS  Lagrangian 
for  the NS-NS background \uga\ is (cf. its RNS 
form in \sffe)
\eqn\ferrt{
L_{F}= i \theta _R \gamma^v \del_+ \theta_R + i \theta_L \gamma^v \del_-
\theta_L 
- 
\four i
 \mm \del_j b_m(x) \  ( \theta_R \gamma^v \gamma^{jm}\theta_R   -
 \theta_L \gamma^v \gamma^{jm}\theta_L)     \ . 
}
Here $b_m$ is any harmonic function.  The background 
 has residual space-time supersymmetry  (so that   \ferr\ has global
fermionic symmetry $\theta \to \theta + \epsilon$) 
provided  $b_m$ is such that 
eq. \qwe\ has non-trivial solutions. 
As we have mentioned in section  2.3, the absence 
of supersymmetries with $\G^u\ep\not=0$
for $\del_i \del_j b_m \not=0$ implies 
that in contrast to the case 
of constant $F_3$ (i.e.  linear $b_m$ \RT)  background 
and the non-constant $F_5$ case in \mam, 
here 
 the \lc gauge GS  model \strr,\ferr\
will  not  have an additional  linearly realized 
2-d supersymmetry.

\medskip

Let us  specify now to the case of the 
R-R background parametrized by an arbitrary 
 holomorphic function, i.e. to the  ``S-dual'' of 
 the $n=1$ background \metra--\eneuno.
According to the discussion in sect. 3.2, this background, 
like its NS-NS counterpart, 
 should be preserving 8 supersymmetries. 
Written in  complex notation, the corresponding 
 light-cone GS Lagrangian  takes the following  form
(cf. \shhh, \ferr,\rekl; here
 we use the notation  $z_1 \equiv z$, $\del_z \eta \equiv  \eta' $)
 \eqn\hhrr{ L_B =  \del_+ z \bd z 
  - \ha \mm ^2 |\eta (z)|^2  
+ \del_+ z_\a \bd z_\a^*\ , 
}
\eqn\hhrm{
L_{F}= i \theta _R \gamma^v \del_+ \theta_R +
i \theta_L \gamma^v \del_-
\theta_L
 - 
\ha i  \mm \ {\rm Re}( \eta' ) \ \theta _L \gamma^v \gamma^{12} \hat P
\theta_R
+
\ha i 
 \mm \ {\rm Im}( \eta' ) \ \theta _L \gamma^v \gamma^{14} \hat P
\theta_R
\ . } 
We have used \caruno, and introduced 
 $$
 \hat P\equiv \ha (1+ \gamma^{1234})\  , 
 $$
which is  the same projector  as  appearing in eq. \soly.
By squaring the Yukawa coupling matrix in \hhrm\ 
it is easy to check that the number of interacting real fermions is 4. 
This does not  match the number (two) of interacting real 
scalars $z=x_1 +ix_3$ -- 
 as expected, the model  we got can not be 2-d supersymmetric.

Explicitly, by choosing an appropriate representation of the
 $\gamma $-matrices 
(see,  e.g.,  section 5.2 in \RT ), 
the  light-cone
gauge  condition $\g^u\theta_{L,R}=0$ can be  solved in terms
 of $8+8$ independent real fermions $S_{L,R}$. Then 
$4+4$ of the fermions will be interacting and  $4+4$ will 
be  massless
(the latter are the ones   which are annihilated by $\hat P$).
By further specifying the $\g$-matrix representation, 
  the
part of the  Lagrangian \hhrr,\hhrm\ describing 
 the system of interacting bosons and fermions can be written as 
$$
L_{int}=\del_+ z \del_- z^* - \ha \mm^2  \eta(z) \eta^*(z^*) 
+  i \psi^*_{kR} \del_+ \psi_{kR} +  i \psi^*_{kL} \del_- \psi_{kL}
$$
 \eqn\horm{
+ \  { i\ov 2}  \mm\ 
\big[  \eta'(z)  \ \psi_{kL}  \psi_{k R}  + \eta'{}^* (z^*) 
 \ \psi^*_{kL}  \psi^*_{k R}  \big] 
\ . } 
Here $\psi_{1\ L,R}$  and $\psi_{2\ L,R }$    are two 
complex combinations  of $4+4$ components of  $S_{L,R}$,
and the sum over $k=1,2$ is implied.\foot{This 
can be shown, e.g., 
by using 4 complex combinations $\hat \g_i$  of $\g$-matrices as in \mam\ 
and writing $S = \psi_i\hat  \g^i \ep_0  + c.c.$, 
where $\ep_0$ is a constant spinor 
satisfying $\hat \g_i \ep_0 =0$.}

This model  is very similar to the (2,2) supersymmetric model 
$L_{(2,2)}= 
\int d^4 \vartheta \ \Phi^* \Phi +  \int d^2 \vartheta \ W(\Phi) + c.c., $
where $ \Phi = z + \vartheta_1 \psi_L  + \vartheta_2 \psi_R + 
...$ and the   superpotential $W(z)$ is  related to an arbitrary  function 
 $\eta(z)$  by 
$W' = \eta$.  
 Indeed, written in  components $L_{(2,2)}$
has the same bosonic part as \horm\ and similar Yukawa terms, 
but just one instead of two pair of complex fermions 
$\psi_L,\psi_R$, i.e. 
$$
L_{(2,2)}=\del_+ z \del_- z^* - \ha \mm^2 \eta(z) \eta^*(z^*) 
+  i \psi^*_{R} \del_+ \psi_{R} +  i \psi^*_{L} \del_- \psi_{L}
$$
 \eqn\hoom{
+ \  {i\ov  \sqrt 2} \mm\ 
\big[  \eta'(z)  \ \psi_{L}  \psi_{ R}  + \eta'{}^* (z^*) 
 \ \psi^*_{L}  \psi^*_{ R}  \big] 
\ . } 
Remarkably, both theories \horm\ and \hoom\ are UV finite:
the mismatch in the number of interacting fermions  is compensated by 
 different coefficients in front of the Yukawa coupling terms.\foot{
 In fact, any model  with one  complex scalar field
 and $K$ species  of fermions (generalising the  $K=1$ \hoom\  and 
 $K=2$ \horm\ models)
 with the Yukawa coupling 
 $ {i\ov  \sqrt {2K}} \mm\ 
\big[  \eta'(z)  \ \psi_{kL}  \psi_{ kR}  + \eta'{}^* (z^*) 
 \ \psi^*_{kL}  \psi^*_{k R}  \big]$ \ ($k=1,...,K$)  
 is also UV finite.
The cancellation of divergences 
 can be readily checked by repeating the argument 
given in section 5.2 (note that only single fermionic loop may 
contribute to the divergences).
As a quick  check of numerical coefficients one may 
consider the case of $\eta(z)=z$  and explicitly integrate out 
$\psi^*_L$ and $\psi^*_R$ in the path integral, getting 
a second-derivative 
action for $\psi_L$ and $\psi_R$, whose contribution cancels 
the bosonic action contribution.}

The (2,2) supersymmetric model \hoom\ was found in \mam\ 
to be a special case of the  \lc  GS action for a particular 
pp-wave background supported  by a nonconstant $F_5$-field. 
The difference between \hoom\ and 
 our model \horm\ corresponding to the $F_3$-background 
is not  unexpected, 
given  
that the  GS  fermionic  couplings in \syym\ 
corresponding to the 
two different exact string solutions -- a background with
 R-R $F_3$ strength  and 
 a background with R-R $F_5$ strength
supporting the same pp-wave metric  -- 
have   different $\G$-matrix structure.
 
While the model \horm\ does not have 2-d supersymmetry,\foot{In particular, 
the vacuum energy on the cylinder does not vanish. In the 1-loop 
approximation: 
$E_1=\sum_{n=1}^\infty \big( 6 n+2 \sqrt{n^2 + M^2} - 
4n- 4  \sqrt{n^2 + \ha M^2}\big)$, 
where $M^2 =\ha \mm^2 |\eta'(z_0)|^2$ \ ($z_0$=const). 
This is to be compared to 
$E_1=  \sum_{n=1}^\infty \big( 6n+2\sqrt{n^2 + M^2} -6n- 
2  \sqrt{n^2 +  M^2}\big)=0$
in the supersymmetric model \hoom.} 
the two models are  very similar, 
representing two different UV finite fermionic 
extensions of the same interacting  bosonic theory with potential 
$V= \mm^2  |\eta(z)|^2$.
For the special choices of $\eta(z)$ for which 
 the bosonic theory \hhrr\ 
and its 2-d supersymmetric version \hoom\ are integrable, 
the same integrability property is likely to be shared also 
by  the theory \horm.

With this motivation in mind we discuss in the next section 
two examples of the holomorphic functions $\eta(z)$ 
which correspond to  integrable models.
As a preparation for the study of string spectrum 
of these models  we shall  make some remarks on the 
solutions of the  Laplace equation in the corresponding
metrics.

\newsec{ Examples 
related  to  integrable models} 

Let us  consider some  examples of  the R-R \lc gauge models 
parametrized by holomorphic functions in the case 
of  the $n=1$ model \hhrr,\hhrm, where
interaction terms depend on 
only  one  complex coordinate field  $z_1\equiv z$.
An  arbitrary holomorphic function $\eta(z)$
which enters \hhrr,\hhrm\ or \horm\ 
can be chosen, as in \mam, to
represent  an integrable 2-d theory.
We shall discuss two
 examples of $\eta(z)$  for which the 
 corresponding (2,2) supersymmetric 
 theories \hoom\ are  the $N=2$ super 
sine-Gordon model \koba\ 
and the $N=2$ super Liouville model \iva.
Since the bosonic parts of \hoom\ and \horm\
are the same, these examples are of interest also 
in the  non-supersymmetric case of the model \horm. 

\subsec{\bf $N=2$ super  Sine-Gordon case }

Choosing  $\eta(z)$ as ($z=x_1+i x_3$,\  $\omega$=real)
\eqn\jio{\eta(z)  =\sin \omega z\ , \ \ \ {\rm i.e.} \ \ \ 
\eta= \sin \om x_1 \ \cosh \om x_3  \ + \ 
i \cos \om x_1 \ \sinh \om x_3 \ , } 
 the bosonic part of the GS Lagrangian
\hhrr,\horm\  becomes the same as the bosonic part of 
the
 $N=2$ super sine-Gordon model  \hoom.

To get some  elementary   information  about 
states of this  theory  which may be useful for future studies,  
 here  we  shall determine 
 the form of the simplest massless  scalar vertex operator
(i.e. the effective masses of  
 scalar states in  the supergravity part of the string spectrum). 
The discussion  below  applies also to the case of 
the  model of \mam\ corresponding to the same pp-wave metric
\gme\  supported not by the $F_3$ but 
by an  $F_5$ background.

Let us consider a   scalar fluctuation $\Psi $
belonging to  the massless supergravity multiplet, which obeys the 
 curved-space Klein-Gordon 
 equation.\foot{
 In general, 
the scalar anomalous dimension operator ($D^2 + ...$)
that  appears in the 
equation for the vertex operator
 may receive $\a'$-corrections (see, e.g., \refs{\CG,\toi} and refs. there).
For example, for the ``null''
NS-NS background in question \gme,\uga\ 
in the bosonic string case  one   would get  corrections like  
 $H^{\mu \a\b} H^{\n}_{\ \a\b} D_\m D_\n= 2(\del_i b_m)^2 \del_v^2$, etc. 
 However, in the superstring case  such corrections are likely to
vanish in an appropriate scheme.
}
Using eq.~\metra , we get 
  \eqn\eww{
  \del_\mu (\sqrt{G} G^{\mu\nu} \del_\nu)  \Psi =
    \big[ 4\del_u\del_v - 4 K(z,z^*) 
\del_v^2 + \del_i^2 \big] \Psi = 0\ .
  }
{}For the $N=2$ sine-Gordon model \jio\ $K$ (which is also 
the bosonic potential in 2-d model) is given by 
\eqn\nnn{
K= -\ha  | \sin (\omega z)|^2=-\four 
\big[ \cosh(2\omega x_3)-\cos(2\omega x_1) \big] \ .}
 The general solution of \eww\ can be 
 obtained as a linear combination of
the waves 
\eqn\hhs{
\Psi = e^{ip_u u+ip_v v+i p_s x_s }\ f(x_1) g(x_3)\ , 
}
where $x_s$ stand for the remaining free coordinates, and 
$f$ and $g$ satisfy  
\eqn\ggss{
f''(x_1)+\big[ a+p_v^2 \cos (2\omega x_1)\big] f=0\ ,\ \ 
}
\eqn\ffoo{ 
 g''(x_3)+\big[e -  p_v^2 \cosh (2\omega x_3) \big]g=0\ ,\
 }
 \eqn\yop{
 4p_u p_v+p^2_s +a+e=0 \ . }
Eq. \ggss\ is the well-known Schr\"odinger 
equation for  the quantum pendulum, and its general solution is expressed
in terms of Mathieu functions.
If $x_1$ is  non-compact 
then  the parameter $a$ takes continuous values.
When $x_1$ is compact
and has period $x_1=x_1+2\pi/\omega $, 
there is a discrete spectrum
of ``Kaluza-Klein" momentum modes.
Since the  vertex operator must be a single-valued function of $x_1$,  
the  Mathieu functions which are solutions of \ggss\ should be 
$2\pi/\omega$  periodic in $ x_1$. This is the case for 
 certain values of $a$ 
-- the Mathieu characteristic eigenvalues -- which thus  
  determine the momentum modes in this sector. 

The second equation \ffoo\  for $g(x_3)$  describes bound states.
Its general solution is also expressed in terms of
Mathieu functions (the associated Mathieu functions of the first kind).
  It is easy  to find the energy eigenvalues using the WKB
approximation.
 The WKB formula gives
 \eqn\goo{
I\equiv  2\int_0^{x_0} dx \sqrt{e-p_v^2\cosh(2 \omega x)} =(n +\ha )\pi
\ , 
  \ \ \ \  \ \
 x_0\equiv {1\over 2\omega }{\rm arccosh}{e\over p_v^2} \ .}
 The integral  is expressed in terms of  an elliptic function,
 \eqn\opp{
 I=-2i {\sqrt{e-p_v^2}\over \omega }  \ 
 {\cal E}( i\omega x_0, -\m)\ ,\ \ \ \ \ \ \ \ 
 \m \equiv {2p_v^2\over e-p_v^2} \ . } 
 {}For $e\gg p_v^2$, we can  approximate \opp\ by 
 \eqn\uio{
 I\approx  {\sqrt{e} \over w} \log {e\over p_v^2 }\approx (n +\ha )\pi \
. } 
 This  determines the eigenvalues $e=e_n$ for large $n$.
 \bigskip
 
The \lc energy, i.e. the value of the \lc Hamiltonian on  
  the corresponding supergravity (i.e. ``massless'' string) 
 states is then given by 
 \eqn\hamp{
 H=-p_u={p^2_s\over 4 p_v}+{1\over 4p_v}\ (a+e_n) \ . 
 }
As was  noted  above, in the case of compact $x_1 = x_1+{2\pi\ov \omega} $,  
the parameter $a$ 
 takes a discrete set of values $a=a_r$
 for which
 the Mathieu equation \ggss\ admits periodic solutions.
 The energy of the physical states in this sector is 
then given by \hamp\ with $a \to a_r$, i.e.
is  parametrized
 in terms of the transverse momentum $p_s$, and two integer numbers
 $r$ and  $n$, i.e. $H=H(p_s,r,n)$.

\subsec{\bf $N=2$ super  Liouville case}

The super Liouville model was
studied in the past, e.g., 
in the context of non-critical string theory   \kuta.
The pp-wave  framework of \mam\ and the present paper 
allows one  to embed a  model with a Liouville potential
into string theory  as a \lc gauge theory 
corresponding  to an exact string solution  with 
 constant dilaton field.

The super Liouville model is obtained by choosing $\eta(z)$ as 
($z= x_1 + i x_3$, \ $\b$=real) 
\eqn\stuu{
\eta(z)  = e^{\beta z}  \ , \ \ {\rm  i.e.}\ \ \ 
\eta= e^{\b x_1} \cos \b x_3 + i e^{\b x_1} \sin \b x_3 \ , 
\ \ \ \ 
 K= -\ha   e^{2\b x_1} \ . } 
One  can  also get  Toda-type potentials 
 by using    more general models
of section 3  with suitably chosen 
holomorphic functions  $\eta_a $.

In the case of the NS-NS background the bosonic part of the 
corresponding covariant   Lagrangian \shhh\ 
takes the following explicit form 
$$
L = \del_+u \bd v -\ha  e^{2\beta x_1}\del_+u \bd u
+\del_+x_1 \bd x_1+\del_+x_3 \bd x_3  
+\del_+ x_s \bd x_s
$$
\eqn\sto{ 
+\  e^{\b x_1}\cos \b x_3 (\del_+ u \bd x_2-
\del_+ x_2 \bd u) -  e^{\b x_1}\sin \b x_3 (\del_+ u \bd x_4-
\del_+ x_4 \bd u)
\ ,  }
where $x_s$ stand for the   remaining free bosonic coordinates.

In the R-R case,  the  light-cone GS Lagrangian 
is given by \hhrr , \hhrm\ (or by \horm)
 \eqn\hhrre{ L_B = \del_+ x_1 \bd x_1 - \ha \mm ^2  e^{{2\beta x_1}}
 + \del_+x_3 \bd x_3 + \del_+ x_s \bd x_s\ , 
}
$$
L_{F}= i \theta _R \gamma^v \del_+ \theta_R + i \theta_L \gamma^v \del_-
\theta_L 
$$
\eqn\hhrme{
- \ 
\ha i \mm \  e^{\b x_1} \cos\b x_3\ \theta _L \gamma^v \gamma^{12} \hat P
\theta_R
+ \ha  i \mm \  e^{\b x_1} \sin\b x_3\  \theta _L \gamma^v \gamma^{14}
\hat P \theta_R\ . 
}
According to the   discussion  in section 5, 
the  fermion couplings here  are different (cf. \hoom\ and \horm)
 from
those of the $N=2$ super Liouville model, which itself can be obtained 
from the pp-wave background with  $F_5$-field of \mam .

As in the sine-Gordon case, 
 one gets two different   models, depending on whether 
 $x_1$ is compact or non-compact.
In the case of compact $x_1$, the semi-classical regime of the Liouville
model
corresponds
to large radius, whereas the quantum regime corresponds to small radius.
In the later  regime, as was pointed out in \mam , 
a more convenient description 
is in terms of  a  mirror theory. 
In the case of the $N=2$ sine-Gordon theory, the mirror
 is a deformed $CP^1$
model, and 
it was argued in \mam\ that the mirror 
 background cannot be a
solution of supergravity as   massive string modes are 
apparently excited.
In the case of the $N=2$ Liouville theory the mirror 
was shown \horkap\ to be
equivalent to the $SL(2,R)/U(1)$ Kazama-Suzuki model, 
which is a supersymmetric  generalization of the 2d black hole model. 
It may be  that in the super Liouville case 
the identification of a 
string background corresponding to  the mirror theory  
is   more direct. 
 
\medskip

Let us now follow section  6.1 and consider  the form of the 
massless scalar  vertex operator in the $N=2$ super Liouville 
theory case.
The vertex operator is again given by eq.~\hhs ,  now with
 $g(x_3)=e^{ip_3 x_3}$, and with $f(x_1)$  determined
by the following differential equation
\eqn\uuoo{
f''+ \big( \nu^2 \b^2 - 2  p_v^2 e^{2\b x_1}\big) f= 0\ ,\ \ \ \ 
\ \ \ \ \
 \nu ^2 \b^2  = - (p^2_s+4 p_u p_v)\ . 
 }
 The  general  normalisable solution is given in terms of the 
 Bessel functions $I_{\nu }$
  \eqn\renyy{
 f(x_1)=  
 i\big[I_{-i\nu }(c\  e^{\b x_1})  - 
 I_{i\nu} ( c \ e^{\b x_1}) \big]\ ,\ \ \ \ c=\sqrt{2}\ {p_v\over\b } \ .
}
The parameter $\nu $ takes continuous real values and 
 represents  momentum of the incoming/outgoing wave
at $x\to -\infty $. 
At $x\to \infty $, the wave is suppressed exponentially due to the
Liouville potential.
The light-cone energy of this state is thus 
\eqn\eenn{
H=-p_u={1\over 4p_v}(p^2_s+\nu^2\b^2) \ . } 

\bigskip

\noindent {\bf Acknowledgements}

\noindent 
We are grateful to J. Maldacena  and R. Metsaev 
for  important  discussions. We also thank 
D. Freedman and  C. Hull for  helpful comments.
 We would like to thank the Newton Institute
for Mathematical Sciences 
and the Aspen Center for Physics for their   hospitality 
while parts of this work were done. 
The work of A.A.T. was supported in part by the grants DOE
DE-FG02-91ER40690, PPARC SPG 00613, INTAS 99-1590
and by the Royal Society Wolfson research merit award.

\vfill\eject \listrefs
\end